\documentclass[sigchi]{acmart}
\usepackage{amsmath,amssymb,amsfonts}
\usepackage{algorithmic}
\usepackage{graphicx}
\usepackage{textcomp}
\usepackage{xcolor}
\usepackage{listings}
\AtBeginDocument{%
  \providecommand\BibTeX{{%
    \normalfont B\kern-0.5em{\scshape i\kern-0.25em b}\kern-0.8em\TeX}}}%
\copyrightyear{2020}
\acmYear{2020}
\setcopyright{acmcopyright}\acmConference[CompSysTech '20]{International
Conference on Computer Systems and Technologies '20}{June 19--20, 2020}{Ruse,
Bulgaria}
\acmBooktitle{International Conference on Computer Systems and Technologies '20
(CompSysTech '20), June 19--20, 2020, Ruse, Bulgaria}
\acmPrice{15.00}
\acmDOI{10.1145/3407982.3408005}
\acmISBN{978-1-4503-7768-3/20/06}
\newtheorem{theorem}{Theorem}
\newtheorem{corollary}{Corollary}
\newcommand{\N}{\mathbb{N}}
\newcommand{\Nplus}{\mathbb{N}^{+}}
\begin{document}
\title[A Method for Fast Computing the Algebraic Degree of Boolean Functions]{A Method for Fast Computing the Algebraic Degree of Boolean Functions}
\author{Valentin Bakoev}
\email{v.bakoev@ts.uni-vt.bg}
\orcid{0000-0003-2503-5325}
\affiliation{
  \institution{``St. Cyril and St. Methodius" University}
  \streetaddress{2 Th. Tarnovski Str.}
  \city{Veliko Tarnovo}
  \state{Bulgaria}
  \postcode{5002}
}
\hyphenation{Boo-le-an}
\begin{abstract}
	The algebraic degree of Boolean functions (or vectorial Boolean functions) is an important cryptographic parameter that should be computed by  fast algorithms. They work in two main ways: (1) by computing the algebraic normal form and then searching the monomial of the highest degree in it, or (2) by examination the algebraic properties of the true table vector of a given function. We have already done four basic steps in the study of the first way, and the second one has been studied by other authors. Here we represent a method for fast computing (the fastest way we know) the algebraic degree of Boolean functions. It is a combination of the most efficient components of these two ways and the corresponding algorithms. The theoretical time complexities of the method are derived in each of the cases when the Boolean function is represented in a byte-wise or in a bitwise manner. They are of the same type $\Theta(n.2^n)$ for a Boolean function of $n$ variables, but they have big differences between the constants in $\Theta$-notation. The theoretical and experimental results shown here demonstrate the advantages of the bitwise approach in computing the algebraic degree---they are dozens of times faster than the byte-wise approaches. 
\end{abstract}

\maketitle
\section{Introduction}
\label{Intro}
  The algebraic degree of Boolean function, or vectorial Boolean function (called S-box), is an important cryptographic parameter. It is used in the design of S-boxes for modern block ciphers, pseudo-random numbers generators in stream ciphers, hash functions, at the Reed-Muller codes, etc. \cite{CC_BFCECC, CC_VBFC, ACANT, CCAGD}. When generating such functions, their cryptographic parameters should be computed by fast algorithms. As faster are these algorithms, more functions will be generated and a better choice among them will be done. At first look, the algorithms for computing the algebraic degree of Boolean functions seem quite simple. Probably because of this, we found only a few papers that consider such algorithms.

	Let $f$ be a Boolean function of $n$ variables given by its Truth Table vector $TT(f)$. The algebraic degree of $f$ can be computed in \textbf{two main ways}. The \textbf{first} one computes the Algebraic Normal Form (ANF) of $f$ and selects the monomial of the highest degree in it. The \textbf{second} way uses only the $TT(f)$, its weight, support, etc., their algebraic properties, and it does not need the ANF of $f$. The algorithms proposed in \cite{CGV} work in this way and use the property: \textit{the algebraic degree of $f$ is maximal iff $TT(f)$ has an odd weight} (its proof is given in \cite{MWNS, CC_BFCECC, ACANT, CGV}). Farther we refer to it as \textit{OddWeight-MaxDeg property}. It is satisfied for half of all Boolean functions and it can be verified very easily and efficiently. When the algorithms in \cite{CGV} are compared with an algorithm of the first type (i.e., based on ANF), some questions arise about the efficiency of algorithms used for computing the ANF and thereafter the algebraic degree. However, we do not know better results concerning the second approach.

%%\hyphenation{Odd-Weight-Max-Deg}
	In \cite{VB-FCAD} we represented our research on the first approach---computing the algebraic degree of Boolean function by its ANF. This paper is its natural continuations in two directions. The first one includes usage of the OddWeight-MaxDeg property, and so it combines the two approaches for computing the algebraic degree. The second one includes improvements outlined in the concluding section of \cite{VB-FCAD}. In this way, we obtain running times that are about 2 times better than these in \cite{VB-FCAD}. These results are represented in this article which is organized as follows. In Section \ref{Bas_Not}, the basic notions are given. In Section \ref{Prelim_Res}, the preliminary results concerning the first way for computing the algebraic degree are represented in short. The method we discuss here is represented in detail in Section \ref{Method}. The experimental results from numerous tests with algorithms of the method are shown in Section \ref{Exp_Res}. The last section contains some concluding notes.
\section{Basic Notions}
\label{Bas_Not}
  Here $\N$ and $\N^{+}=\N \backslash \{0\}$ are the set of natural numbers and the set of positive natural numbers, correspondingly. The set of all $n$-dimensional binary vectors is known as \textit{$n$-dimensional Boolean cube} and it is defined as $\{0,1\}^n=\{(x_1$, $x_2,\dots,x_n)|$ $x_i\in \{0,1\}, \forall\, i=1,2,\dots, n \}$. So $|\{0,1\}^n| =|\{0,1\}|^n= 2^n$. 

	For an arbitrary vector $\alpha=(a_1, a_2, \dots, a_n)\in \{0,1\}^n$, 
the natural number $\#\alpha=\sum_{i=1}^n a_i.2^{n-i}$ is called a  \textit{serial number} of the vector $\alpha$. So, the $n$-digit binary representation of $\#\alpha$ is $a_1 a_2 \dots a_n$. A \textit{(Hamming) weight} of $\alpha$ is the natural number $wt(\alpha)$, equal to the number of non-zero coordinates of $\alpha$, i.e., $wt(\alpha)= \sum_{i=1}^n a_i$. For any $k\in \N$, $k\leq n$, the set of all $n$-dimensional binary vectors of weight $k$ is called a \textit{$k$-th layer} of the $n$-dimensional Boolean cube. It is denoted by $L_{n,k}= \{\alpha|\, \alpha \in \{0,1\}^n : wt(\alpha)=k\}$ and we have $|L_{n,k}|= \binom{n}{k}$, for $k=0, 1, \dots, n$. The family of all layers $L_n=\{ L_{n,0}, L_{n,1},\dots,L_{n,n}\}$ is a \textit{partition} of the $n$-dimensional Boolean cube into layers and so $\sum_{k=0}^n \binom{n}{k}= 2^n= |\{0, 1\}^n|$. 

	For arbitrary vectors $\alpha= (a_1, a_2, \dots, a_n)$ and $\beta= (b_1, b_2,$ $\dots, b_n) \in \{0,1\}^n$, we say that ``$\alpha$ \textit{precedes lexicographically} $\beta$" and denote it by $\alpha\leq \beta$, if $\alpha=\beta$ or if $\exists\, k, 1\leq k\leq n$, such that $a_k<b_k$ and $a_i=b_i$, for all $i<k$. The relation ``$\leq$" defines a \textit{total} order in $\{0,1\}^n$, called \textit{lexicographic order}. The vectors of $\{0,1\}^n$ are ordered lexicographically in the sequence $\alpha_0, \alpha_1, \dots  \alpha_{2^n-1}$ if and only if the sequence of their serial numbers is  $0, 1, \dots, 2^n-1$.
	
	A \textit{Boolean function} of $n$ variables is a mapping $f:\{0,1\}^n  \rightarrow \{0,1\}$. So, if $x_1, x_2, \dots , x_n$ denote the variables of $f$, then $f$ maps any binary input $x=(x_1, x_2,\dots , x_n) \in \{0,1\}^n$ to a single binary output $y = f(x)\in \{0,1\}$. Any Boolean function $f$ can be represented in a unique way by the vector of its functional values, called a \textit{Truth Table} vector and denoted by $TT(f)= (f_0, f_1, \dots f_{2^n-1})$, where $f_i= f(\alpha_i)$ and $\alpha_i$ is the $i$-th vector in the lexicographic order of $\{0,1\}^n$, for $i= 0, 1, \dots, 2^n-1$.	The set of all Boolean functions of $n$ variables is denoted by ${\mathcal{B}}_n$ and its size is $|{\mathcal{B}}_n|= 2^{2^n}$.

	Another unique representation of the Boolean function $f\in {\mathcal{B}}_n$ is the \textit{algebraic normal form} (ANF) of $f$, which is a multivariate polynomial
\begin{eqnarray*}
\label{F10}
f(x_1, x_2, \dots, x_n)= \bigoplus_{\gamma\in \{0,1\}^n} a_{\#\gamma}\, x^{\gamma}\,.
\end{eqnarray*}
	
	Here $\gamma=(c_1, c_2, \dots, c_n)\in \{0,1\}^n$, the coefficient $a_{\#\gamma}\in \{0, 1\}$, and $x^{\gamma}$ means the monomial $x_1^{c_1} x_2^{c_2} \dots x_n^{c_n}= \prod_{i=1}^n x_i^{c_i}$, where $x_i^0=1$ and $x_i^1=x_i$, for $i=1, 2, \dots n$. A \textit{degree} of the monomial $x=x_1^{c_1} x_2^{c_2} \dots x_n^{c_n}$ is the integer $deg(x)= wt(\gamma)$ which is the number of variables of the type $x_i^1=x_i$. The \textit{algebraic degree} (or simply \textit{degree}) of $f$ is defined as $deg(f)= max\{deg(x^{\gamma})|\, a_{\#\gamma}=1\}$. 
	
	When $f\in {\mathcal{B}}_n$ and the $TT(f)$ is given, the  coefficients $a_0, a_1, \dots, a_{2^n-1}$ can be computed by a fast algorithm, called an \textbf{\textit{ANF transform}} (ANFT)\footnote{In dependence of the area of consideration, the same algorithm is called also (fast) M\"obius Transform, Zhegalkin Transform, Positive polarity Reed-Muller Transform, etc.}. The ANFT is well studied, it is derived in different ways by many authors, for example \cite{CC_BFCECC, ACANT, AJOUX}. Its byte-wise implementation has a time-complexity $\Theta(n.2^n)$. The vector $A_f=(a_0, a_1, \dots, a_{2^n-1})\in \{0,1\}^n$ denotes the result obtained after the ANFT. When $f\in {\mathcal{B}}_n$ is the constant zero function (i.e., $TT(f)=(0,0,\dots, 0)$), its ANF is $A_f= (0,0,\dots,0)$ and its algebraic degree is defined as $deg(f)= -\infty$. If $f$ is the constant one function ($TT(f)=(1,1,\dots, 1)$), then $A_f= (1,0,0,\dots,0)$ and $deg(f)=0$. 
\section{Preliminary Results}
\label{Prelim_Res}
	We shall represent in short our preliminary investigations and results which are focused in 4 directions. They concern only the first approach---computing $deg(f)$ after computing the vector $A_f$ of arbitrary $f\in {\mathcal{B}}_n$ given by its $TT(f)$. By using bitwise data representation and bitwise operations, several efficient algorithms for computing $deg(f)$ were developed.
\subsection{Bitwise Implementation of the ANFT}
\label{BitwiseANFT}
	A comprehensive study of the bitwise implementation of the ANFT is proposed in \cite{VB-FastANFT}. This version of the algorithm uses 64-bit computer words and then its time complexity is $\Theta((9n-2).2^{n-7})$, and the space complexity is $\Theta(2^{n-6})$. As we noted above, the byte-wise version of the ANFT algorithm has a time-complexity $\Theta(n.2^n)$, i.e, of the same type. However, the experimental results show that the bitwise version of the algorithm is about 25 times faster in comparison to the byte-wise version (for Boolean functions of $5, 6, 8, 10, 12, 16$ variables and at the parameters of the tests used in \cite{VB-FastANFT}). In \cite{DBIB}, analogous research of the parallel bitwise implementation of the ANFT has been done and similar results have been obtained for its efficiency.
\subsection{Distribution of Boolean Functions According to their Algebraic Degrees}
\label{Distrib}
\hyphenation{su-re-ly}
	The OddWeight-MaxDeg property is well known, as it was noted above. But not so popular is the property ``When $n$ tends to infinity, random Boolean functions have almost surely algebraic degrees at least $n-1$." \cite[p. 49]{CC_BFCECC}. We consider that the complete enumeration and distribution of all $f \in {\mathcal{B}}_n,  n\in \Nplus$, according to their algebraic degrees is very important for our research. Their study was done in \cite{VB-DISTR}. Some of the main results in it concern the number $d(n,k)$ of all Boolean functions $f \in {\mathcal{B}}_n : deg(f)= k$.
\begin{theorem}
\label{Th10} 
	For any integers $n\geq 0$ and $0\leq k\leq n$, the number
\begin{eqnarray*}
\label{F20}	
d(n, k)=\left\{
\begin{array}{ll}
	 1, & \textrm{\ if\ } k=0;\\
	 (2^{\binom{n}{k}}-1).2^{\sum_{i=0}^{k-1} \binom{n}{i}}, & \textrm{\ if\ } 1\leq k\leq n.\\
\end{array}
\right.
\end{eqnarray*}
\end{theorem}
\begin{corollary}
\label{Cor10}
	The number $d(n,n-1)$ tends to $\displaystyle \frac{1}{2}\cdot|{\mathcal{B}}_n|$ when $n\rightarrow\infty$.
\end{corollary}

	The main conclusion from these assertions is: when $n \to \infty$, almost all $f \in {\mathcal{B}}_n$ have an algebraic degree $n$ or $n-1$. It and the OddWeight-MaxDeg property have crucial meaning when creating our algorithms. 
\subsection{Weight-Lexicographic Order of the Vectors of $\{0,1\}^n$}
\label{WLO}
	The simplest algorithm for computing the algebraic degree of an arbitrary $f\in{\mathcal{B}}_n$ is an \textit{Exhaustive Search} (refered as \textbf{\textit{ES algorithm}}): when $A_f=(a_0, a_1$, $\dots, a_{2^n-1})$ is given, it checks consecutively whether $a_i=1$, for $i=0,1,\dots, 2^n-1$. The algorithm selects the vector of maximal weight among all vectors $\alpha_i\in \{0,1\}^n$ such that $a_i=1$. Since the ES algorithm checks exhaustively all values in $A_f$ (corresponding to the lexicographic order of the vectors of $\{0,1\}^n$), it performs $\Theta(2^n)$ checks. However, if we use an algorithm that checks the values of $A_f$ according to the vectors' weights, it will finish after the first check for the half of all $f \in {\mathcal{B}}_n$, or after no more than $n$ checks for almost all of the remaining Boolean functions---according to the main conclusion drawn above. The main parts of  faster ways for computing the algebraic degree are proposed in \cite{VB-WLO-A} where two important orders of the vectors of $\{0,1\}^n$ are  studied. So, the \textit{sequence of layers} $L_{n,0}, L_{n,1}, \dots, L_{n,n}$ defines an \textit{order} of the vectors of $\{0,1\}^n$ according to their weights. The corresponding relation $R_{<_{wt}}$ is defined as follows: for arbitrary $\alpha, \beta \in \{0, 1\}^n$, $(\alpha, \beta) \in R_{<_{wt}}$ if $wt(\alpha) < wt(\beta)$ or if $\alpha=\beta$. Then we say that "$\alpha$ \textit{precedes by weight} $\beta$" and write also $\alpha <_{wt} \beta$. So, $R_{<_{wt}}$ is a partial order in $\{0, 1\}^n$ and we call it (and the order determined by it) a \textit{Weight-Order} (WO).
	
	For an arbitrary layer $L_{n,k}=\{\alpha_0, \alpha_1, \dots, \alpha_m\}$ of $\{0,1\}^n$, the \textit{sequence of serial numbers} of the vectors of $L_{n,k}$ is defined as $l_{n,k}=\#\alpha_0, \#\alpha_1, \dots, \#\alpha_m$. If $l_n= l_{n,0}, l_{n,1}, \dots, l_{n,n}$ is the \textit{sequence of all serial numbers} corresponding to the vectors in the sequence of layers $L_{n,0}, L_{n,1}, \dots, L_{n,n}$, then $l_n$ represents a WO of the vectors of $\{0,1\}^n$. We call $l_n$ a \textit{WO sequence} of $\{0,1\}^n$. One of all possible $\prod_{k=0}^n \binom{n}{k}!$ WO sequences deserves a special attention. It can be obtained by two algorithms: the first one is similar to the known Bucket sort algorithm \cite{CLRS}, whereas the second algorithm is created by using a special definition. In both ways, a \textit{total weight order} is obtained for the sequence $l_n$, where the lexicographic order is a second criterion for ordering the vectors of equal weights. It is called a \textit{Weight-Lexicographic Order} (WLO).
\begin{table}[!ht]
	\caption{WLO sequence $l_n$, for $n=1,2,3,4$}
	\label{Tab:Results-WLO-Alg}
	\small{
	\begin{center}
		\begin{tabular}{c|l}
		\hline
		$ n $ &\, $l_n$\\
		\hline
		1  &\, 0, 1 \\
		\hline
    2  &\, 0, 1, 2, 3 \\
    \hline
    3  &\, 0, 1, 2, 4, 3, 5, 6, 7 \\
    \hline
    4  &\, 0, 1, 2, 4, 8, 3, 5, 6, 9, 10, 12, 7, 11, 13, 14, 15 \\
		\hline
		\end{tabular}
	\end{center}
	}
\end{table}

	Another way for representation of the vectors from the layer $L_{n,k}$ is by the \textit{characteristic vector} $m_{n,k}$ \textit{of the layer} $L_{n,k}$, for $k=0,$ $1,\dots, n$. It is defined as follows: $m_{n,k}= (c_0,c_1, \dots, c_{2^n-1})\in \{0,1\}^{2^n}$, where:
\begin{eqnarray*}
c_i=\left\{
\begin{array} {ll}
0, \textrm{\ if\ } \alpha_i\notin L_{n,k}\,,\\
1, \textrm{\ if\ } \alpha_i\in L_{n,k}\,,
\end{array}
\right.
\end{eqnarray*} 	
$\alpha_i\in \{0,1\}^n$, for $i=0, 1, \dots ,2^n-1$.	

	The sequence of vectors $m_{n,k}$, for $k= 0,1,\dots, n$, can be obtained by two algorithms again. The first one put units in all bits of $m_{n,k}$ whose coordinates are determined by the members of $l_{n,k}$, for $k= 0,1,\dots, n,$. The second algorithm uses a special definition. These vectors are used as \textit{masks} (the notation $m$ comes from mask) when the algebraic degree is computed by a bitwise algorithm. It determines $deg(f)$ by computing the conjunction $A_f\wedge m_{n,k}$, for $k=n, n-1, \dots, 0$, until $A_f\wedge m_{n,k}=0$. The first value of $k$ such that $A_f\wedge m_{n,k}>0$ means that $deg(f)=k$.
\subsection{Fast Computing the Algebraic Degree of Boolean Functions}
\label{Fast_Comp_AD}
	In \cite{VB-FCAD} we consider two approaches in computing the algebraic degree: a \textit{byte-wise} and a \textit{bitwise}. The first approach includes ES algorithm and the so called \textit{Byte-wise WLO algorithm}. Let $f \in {\mathcal{B}}_n$ be given by the vector $TT(f)$. Suppose that $l_n=(i_0, i_1, \dots, i_{2^n-1})$ which is a permutation of all integers between $0$ and $2^n-1$, and $A_f=(a_0, a_1, \dots, a_{2^n-1})$ are already computed. The Byte-wise WLO algorithm checks the values of $A_f$ by using the members of the sequence $l_n$, from the last to the first one. So, for $j= i_{2^n-1}, i_{2^n-2},\dots, 2,1,0$, it checks consecutively the $j$-th coordinate of $A_f$, until it is $=0$. If all coordinates of $A_f$ are $=0$, then $f$ is the constant zero function. Otherwise, if $i_j$ is the first coordinate of $A_f$ which is $=1$ and if $i_j$ is a member of the subsequence $l_{n,k}$, then the algorithm stops and returns $k$ since $k=deg(f)$. The time complexity of this algorithm is $O(2^n)$ in the general case, although the algorithm makes no more than $n+1$ checks at almost all $f\in{\mathcal{B}}_n$, especially when $n$ grows.

	A representative of the bitwise approach is the \textbf{\textit{Bitwise WLO algorithm}}. We accept that it always uses masks. The idea of how it works was just given above. When $A_f$ occupies one computer word, the algorithm uses $n+1$ masks. So it performs at most $n+1$ steps and its time complexity is $O(n)$, which is of logarithmic type ($n=\log_2{2^n}$) with respect to the input size. When the size of computer word is $64=2^6$ bits and $f$ is a function of $n>6$ variables, $A_f$ occupies $s=2^{n-6}$ computer words. So, $m_{n,k}$ will occupy $s$ computer words also and the computing $A_f \wedge m_{n,k}$ will be done in $s$ steps, for $k=n, n-1, \dots, 0$. If on some of these steps the conjunction between the corresponding computer words of $A_f$ and $m_{n,k}$ is greater than zero, the algorithm returns $k$ and stops. So, the general time complexity of the algorithm is $O(n).O(s)=O(n.2^{n-6})$. For details, the reader may refer to Example 1 in \cite{VB-FCAD} which demonstrates how the WLO algorithms work. In the following C/C++ code of the Bitwise WLO algorithm, the masks are represented by a two-dimensional array. The number of its rows is \verb#n_vars + 1# (where \verb#n_vars# is the number of variables), and the number of columns is equal to the number of computer words (\verb#n_cwords#) used for the representation of $A_f$.
{\small
\begin{lstlisting}[language=C, caption=Bitwise WLO algorithm]
typedef unsigned long long ull;
int max_deg_by_masks (ull A_f[]) {
  for (int row= n_vars; row >= 0; row--) {
    for (int col= 0; col < n_cwords; col++){	
      if (A_f[col] & masks[row][col])			
        return row; // the layer's number
    }               // which is =deg(f)
  }
  return -1; // when f is the 0-constant
}
\end{lstlisting}
}
	Numerous tests have been conducted to compare the efficiency of algorithms considered. A test file of $10^8$ randomly generated unsigned integers (in 64-bit computer words) has been used as input, i.e., as $TT(f)$-s.  Depending on the number of variables $n$, the serial function to be tested is formed by reading $s$ consecutive integers from the file and so $10^8/s$ such functions are tested. The algebraic degrees of Boolean functions of $n=5, 6, 8, 10, 12$ and $16$ variables have been computed after computing the $A_f$ of the corresponding functions.  The experimental results show that the Bitwise WLO algorithm is 20 and more times faster than the Byte-wise algorithms. The conclusions after these results were used for extensions and improvements of the algorithms, as well as the tests' parameters.
\section{A Method for Fast Computing the Algebraic Degree}
\label{Method}
	The continuation of our research involves the OddWeight-MaxDeg property.  When $f\in{\mathcal{B}}_n$, the weight of $TT(f)$ can be computed efficiently. Since $wt(TT(f))$ is odd for half $f\in{\mathcal{B}}_n$, they all have an algebraic degree $=n$. So this property is part of the algorithms proposed in \cite{CGV} which represent the second way of computing the algebraic degree. But using the same algorithms for the remaining half $f\in{\mathcal{B}}_n$ is not efficient enough---this conclusion follows from the analysis of the numerical results (in Sect. 4 of \cite{CGV}) and after comparing them with the results from \cite{VB-FCAD}. That is why for this half of Boolean functions we will use the first way of computing the algebraic degree and thus we combine both ways. In addition, and as we have shown in Section \ref{Prelim_Res}, the first way includes several algorithms of different types, their mathematical bases are also different \cite{VB-FastANFT, VB-DISTR, VB-WLO-A, VB-FCAD}. These are the reasons to talk about a \textbf{\textit{method}} that integrates these algorithms instead of just talking about algorithms.
	
	The usage of OddWeight-MaxDeg property in our method means efficient computing of $wt(TT(f))$. Such algorithms are considered in \cite{RND}. When $f\in {\mathcal{B}}_n$ is represented by its $TT(f)$ in a byte-wise manner, the computing of its weight needs $\Theta(2^n)$ operations. When $TT(f)$ is represented in $s= 2^{n-6}$ computer words (of $64$ bits), $wt(TT(f))$ can be computed efficiently by $\Theta(6.2^{n-6})$ operations (as it is shown in \cite{RND}), or by $\Theta(4.2^{n-6})$ operations if a look-up table\footnote{An array $w$ where the weight of integer $i$ is precomputed and stored in $w[i]$, for $i=0, 1,\dots, 65535$, as it is shown in \cite{IBVB, VB-WLO-A}.} of size $2^{16}$ is used. However, the exact value of the weight is not necessary, it is only important to know whether it is an odd number or not. This can be achieved by an algorithm that computes a \textit{Parity Check} (PC) of $TT(f)$. We developed such an algorithm, referred to as a bitwise \textbf{\textit{PC algorithm}}. Its input is the vector $TT(f)$ represented by the array \verb#TT_f# of $s$ unsigned $64$-bit integers (\verb#n_cwords# represents $s$ again). Here is its C/C++ code.
{\small
\begin{lstlisting}[language=C, caption=Bitwise PC algorithm]
int parity_check (ull TT_f[]) {
  int sum= TT_f[0];
  for (int i= 1; i < n_cwords; i++) {
    sum ^= TT_f[i];
  }
  for (int i= 32; i > 0; i >>= 1) {
    sum ^= (sum >> i);
  }
  return (sum & 1); // 1 if wt(TT(f)) is odd
}
\end{lstlisting}
}		
	 Firstly, the bitwise PC algorithm performs bitwise XORs (sums modulo 2) between all elements of the array \verb#TT_f# and stores the result in the variable \verb#sum#. Thus, the leftmost bit of \verb#sum# is a result of sum modulo 2 between the leftmost bits of all elements of \verb#TT_f#, the second bit of \verb#sum# is a result of sum modulo 2 between the second bits of all elements of \verb#TT_f#, etc. So, after $s-1$ bitwise XORs, the bits of \verb#sum# contain the result of all XORs between the corresponding bits of the elements of \verb#TT_f#. Secondly (by the second cycle), the bitwise PC algorithm continues with sum modulo 2 between the left and right half of \verb#sum#, i.e., \verb#sum ^= (sum >> 32)#. Let the bits of \verb#sum# be numbered by $0,1,\dots, 63$, from left to the right. Thus the right half of \verb#sum# contains the result of all XORs so far, as it is obtained by XORs between the corresponding bits---these with numbers: 0 and \textbf{32}; 1 and \textbf{33}; etc., 31 and \textbf{63} where the bits of bold numbers contain the result. The next step is \verb#sum ^= (sum >> 16)# because we are interested in the result between the corresponding bits in the  the rightmost 2 (i.e., third and fourth) quarters of \verb#sum#---these with numbers: 32 and \textbf{48}; 33 and \textbf{49}; etc., 47 and \textbf{63}. Analogously, the PC algorithm continues with \verb#sum ^= (sum >> 8)#, etc., \verb#sum ^= (sum >> 1)#. After the last XOR the rightmost bit of \verb#sum# is the \textbf{\textit{parity check bit}}---its value is 1 if $TT(f)$ has an odd weight, otherwise it is 0. So, the PC algorithm executes $6$ additional steps and its total time complexity is $\Theta(s-1+6)= \Theta(2^{n-6})$. Thus, it should be faster than the algorithms for computing the weight of $TT(f)$ and the experimental results confirm this theoretical conclusion. We note the correctness of the PC algorithm follows from the associative property of the XOR operation, the appropriate choice of pairs of bits that should be XOR-ed and where (in which bits) the result is stored. The correctness can be rigorously proven by using these arguments.
	
	Let $f\in {\mathcal{B}}_n$ is represented by its $TT(f)$ in a \textbf{bitwise} manner. In this case, the \textbf{method} for fast computing the algebraic degree of $f$ works as follows.
\\
\textbf{Step 1}. Use the PC algorithm to compute the PC bit of $TT(f)$. If its value is $1$, then return $n$ ($=deg(f)$) and stop.
\\
\textbf{Step 2}. Use the bitwise ANFT algorithm and compute $A_f$.
\\
\textbf{Step 3}. Use the Bitwise WLO algorithm to compute $deg(f)$.

%% \hyphenation{al-go-rithm}
	In \cite{VB-FCAD} we outlined a new idea for another bitwise algo-rithm---to check the bits of $A_f$ according to the WLO sequence. So it will be analogous to the byte-wise WLO algorithm and it will have the same time complexity $O(2^n)$. We first ignored this idea, since the time complexity of the bitwise WLO algorithm is $O(n.2^{n-6})$ and $n.2^{n-6} < 2^n$ when $6\leq n<64$. But after that, we realized that for almost $100\%$ of all $f\in {\mathcal{B}}_n$, the bitwise WLO algorithm performs $O(2^{n-6})$ checks, whereas the byte-wise WLO algorithm (as well as the new bitwise algorithm) performs $O(n)$ checks. In addition, checking the serial bit of $A_f$ in accordance with the WLO sequence requires a maximum of 5 operations. So, the new bitwise algorithm will have a small constant hidden in the $O$-notation. We call it \textit{Check Bits in WLO algorithm}, or shortly \textbf{\textit{CB WLO algorithm}}. Obviously, the bitwise WLO algorithm will be faster than the CB WLO algorithm for a small $n$ (say $n\leq 8$), but for $n=16$ it will perform much more operations than the CB WLO algorithm. Thus, the CB WLO algorithm can be used instead of the Bitwise WLO algorithm in Step 3 of the method discussed. The results of such a replacement can be seen in the next section.

  The total time complexity of the method is a sum of the time complexities of the algorithms in its steps---for the bitwise version in the general case we have: $\Theta(2^{n-6}) + \Theta((9n-2).2^{n-7}) + O(2^n)= \Theta(n.2^n)$.
	
  When $f\in {\mathcal{B}}_n$ is represented by its $TT(f)$ in a \textbf{byte-wise} manner, the \textbf{method} works with corresponding byte-wise algorithms.
\\
\textbf{Step 1}. Use a byte-wise algorithm that computes $wt(TT(f))$. If it is an odd number, then return $n$ and stop.
\\
\textbf{Step 2}. Use the byte-wise ANFT algorithm and compute $A_f$.
\\
\textbf{Step 3}. Use the byte-wise WLO algorithm to compute $deg(f)$.  
  
  In this case, the total time complexity of the method is: $\Theta(2^n) + \Theta(n.2^n) + O(2^n)= \Theta(n.2^n)$. So, in both cases we obtained time complexities of the same type. The differences between them are in the constants hidden in the $\Theta$-notation.
\section{Experimental Results}
\label{Exp_Res}
	We have conducted a lot of tests to verify and understand what these theoretical time complexities mean in practice. To obtain more precise results in comparison with these in \cite{VB-FCAD}, we have done some \textbf{essential changes} in the tests' parameters, methodology of testing, etc. So, we used the largest test file (of size $\approx 14$ GB) which contains $10^9$ randomly generated unsigned integers, each in a single 64-bit computer word. The file was checked for representativeness, as shown in \cite{VB-DISTR}. Other  important tests' parameters are:
\begin{enumerate}
	\item \textit{Hardware parameters}: Intel Pentium CPU G4400, 3.3 GHz; the RAM was enlarged to 16 GB, so that the whole test file can be read and located in it. 
	\item \textit{Software parameters}: Windows 10 OS and MVS Express 2015 for Windows Desktop. The algorithms are written in C++, in a single program which is built in Release mode as 64-bit console application (when the program is built as 32-bit console application, it runs slower). 
	\item \textit{Methodology of testing}: all tests were executed 5 times, on the same computer, under the same conditions, and without an Internet connection. The smallest and the biggest running time are ignored, and the remaining 3 running times are taken on average. All results were checked for coincidence. The time for conversion to byte-wise representation is excluded from all running times in the following tables.
\end{enumerate}

	Improvements in the testing methodology have resulted in better and more reliable results. For example, the maximum difference between the three test results and their average is less than $0.01\%$ (in the 72 basic tests and 54 intermediate tests). There are only 5 exceptions---for 5 tests this value is less than $0.4\%$.
	
\begin{figure}[ht]
   \centering
       \scalebox{0.66}{\includegraphics{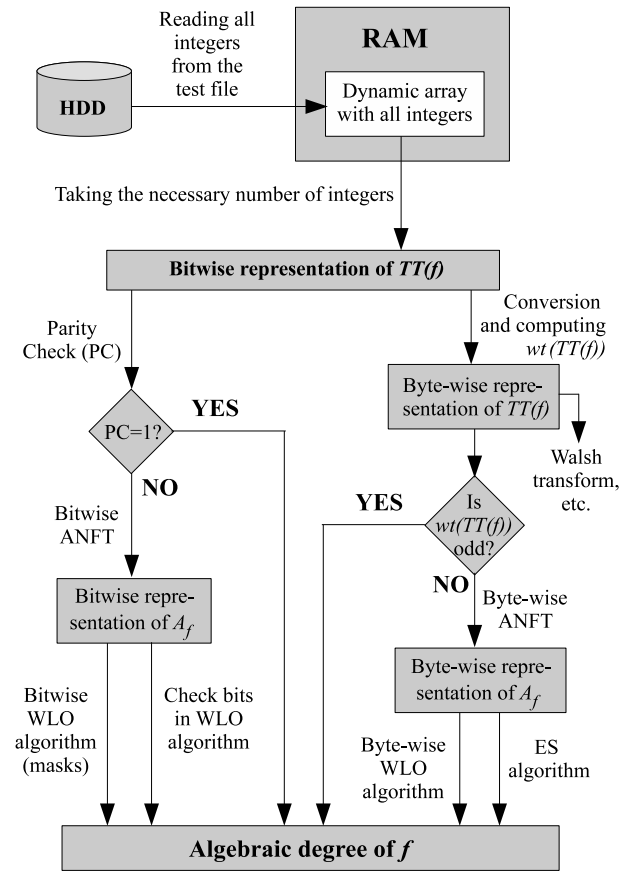}}
   \caption{Scheme of performance of the tests}
   \label{Fig:Scheme-New}    
\end{figure}

	A scheme of the computations and used algorithms in the two versions of the method is shown in Fig. \ref{Fig:Scheme-New}. If we compare it with the corresponding Figure 1 in \cite{VB-FCAD} we will notice the evolution and improvements we discuss here. When Boolean functions (BFs) of 6 and more variables are tested, $2^{n-6}$ consecutive integers are taken from the dynamic array and so they form the serial Boolean function. The results from the tests are shown in Table \ref{Ta2:ByteWiseAlgorithms} and Table \ref{Ta3:BitWiseAlgorithms} so that the byte-wise and bitwise versions can be easily compared. On the other hand, two more cases are tested and shown in the tables: (1) when PC (or computing the weight) is not performed and (2) when it is performed before the execution of the remaining algorithms. In this way, we can compare the running times when the algebraic degree is computed by the algorithms from \cite{VB-FCAD} and when it is computed by the method under consideration. 
\begin{table*}[!ht]
	\caption{Experimental results for the byte-wise algorithms}
  \label{Ta2:ByteWiseAlgorithms}
  \centering
  \small{
  \begin{tabular}{|c|c|c|c|c|c|c|c|c|c|}
  \hline
  Byte-wise & \multicolumn{9}{c |}{Pure running time in seconds for Boolean functions of:} \\
		  \cline{2-10} 
    implementation  &	6 vars,    & 7 vars,      & 8 vars,      & 10 vars,     & 11 vars,      & 12 vars,      & 14 vars,       & 15 vars        & 16 vars, \\
    of:         & \footnotesize{$10^9$ BFs} & \footnotesize{$10^9/2$ BFs} & \footnotesize{$10^9/4$ BFs} & \footnotesize{$10^9/2^4$ BFs} & \footnotesize{$10^9/2^5$ BFs} & \footnotesize{$10^9/2^6$ BFs}  & \footnotesize{$10^9/2^8$ BFs} & \footnotesize{$10^9/2^9$ BFs} & \footnotesize{$976\,562$ BFs} \\  
  \hline
  ANFT+ES  & 343.278 & 350.640 & 355.627 & 355.744 & 370.639 & 397.470 
 & 402.885 & 400.680 & 406.506 \\
  \hline
  ANFT+WLO & 158.571 & 142.279 & 141.791 & 123.915 & 124.867 & 142.886 & 149.825 & 146.783 &  149.816 \\
  \hline 
  PC+ANFT+ES  & 180.389 & 183.237 & 184.748 & 174.291 & 190.840 & 209.423 & 199.476 & 197.720 & 206.412 \\
  \hline
  PC+ANFT+WLO &  83.969 &  75.004 &  70.456 &  45.610 &  47.086 &  65.336 &  71.970 & 71.052 &  76.902 \\
  \hline
	\end{tabular}
}
\end{table*}

\begin{table*}[!ht]
	\caption{Experimental results for the bitwise algorithms}
  \label{Ta3:BitWiseAlgorithms}
  \centering
  \small{
  \begin{tabular}{|c|c|c|c|c|c|c|c|c|c|}
  \hline
  Bitwise & \multicolumn{9}{c |}{Pure running time in seconds for Boolean functions of:} \\
		  \cline{2-10} 
    implementation  &	6 vars,    & 7 vars,      & 8 vars,      & 10 vars,     & 11 vars,     &  12 vars,      & 14 vars,       & 15 vars        & 16 vars, \\
    of:        & \footnotesize{$10^9$ BFs} & \footnotesize{$10^9/2$ BFs} & \footnotesize{$10^9/4$ BFs} & \footnotesize{$10^9/2^4$ BFs} & \footnotesize{$10^9/2^5$ BFs} & \footnotesize{$10^9/2^6$ BFs}  & \footnotesize{$10^9/2^8$ BFs} & \footnotesize{$10^9/2^9$ BFs} & \footnotesize{$976\,562$ BFs} \\  
  \hline
  ANFT+WLO  & 10.954 & 7.916 & 6.852 & 6.301 & 6.327 & 6.393 & 6.928 & 6.940 & 6.596 \\
  \hline 
  ANFT+CB WLO & 16.632 & 10.928 & 8.197 & 5.896 & 5.667 & 5.662 & 6.016 & 6.145 & 6.161 \\
  \hline  
  PC+ANFT+WLO  & 10.373 & 7.094 & 5.184 & 4.096 & 3.673 & 3.493 & 3.623 & 3.671 & 3.668 \\
  \hline
  PC+ANFT+CB WLO & 16.005 & 9.891 & 6.520 & 4.124 & 3.616 & 3.402 & 3.507 & 3.559 &  3.563 \\
  \hline
	\end{tabular}
}
\end{table*}
\section{Conclusions}
\hyphenation{me-thod}
	Here we represented and discussed a method for fast computing the algebraic degree of Boolean functions and its byte-wise and bitwise versions. For each of them, we derived its total time complexity. The experimental results confirmed the theoretical conclusions---the bitwise version of the method (or algorithms) is dozens of times faster than the corresponding byte-wise version, especially when $n$ grows. Some other conclusions are:
\begin{itemize}
	\item The total running time only of ANFT + ES algorithms grows simultaneously with the growth of $n$. But this is not true for the total running times of the remaining byte-wise algorithms, they have local minimums when $n=10$. For all bitwise algorithms, the growth of $n$ does not correspond to the growth of their total running times---there are local minimums when $n=10$, for ANFT + WLO algorithms, and when $n=12$, for all remaining algorithms. These interesting facts need to be analyzed and explained.
	\item There are intermediate experimental results which are not shown in the tables. For example, the conversion from byte-wise to a bitwise representation of $TT(f)$ is a very expensive operation, its running time is more than 2 times higher than the total running time of PC + ANFT + WLO (byte-wise) algorithms. That is why the byte-wise version of the method should be used when the Boolean functions are represented in a byte-wise manner at the input---for example, when they are generated by such algorithms.
\end{itemize}

  Probably some improvements to the algorithms in the method will be achieved by the processor instructions for simulating 128-bit, 256-bit or more bit arithmetic. Such experiments are forthcoming.
\section*{Acknowledgment}
  This work was partially supported by the Research Fund of the University of Veliko Tarnovo (Bulgaria) under contract No FSD-31-299-05/05.05.2020.
\end{document}